\documentclass[runningheads]{llncs}

\usepackage{booktabs}   
\usepackage{subcaption} 
                        
\usepackage[utf8]{inputenc} 
\usepackage[T1]{fontenc}    
\usepackage{hyperref}       
\usepackage{url}            
\usepackage{booktabs}       
\usepackage{amsfonts}       
\usepackage{nicefrac}       
\usepackage{microtype}      
\usepackage{xcolor}
\usepackage{graphicx}
\usepackage{multirow}
\usepackage{cite}
\usepackage{appendix}
\usepackage{listings}
\usepackage{amsmath}

\usepackage{algorithm}
\usepackage{algorithmicx}
\usepackage{algpseudocode}
\MakeRobust{\Call}
\floatname{algorithm}{Algorithm}

\begin{document}

\title{Auto-Precision Scaling for Distributed Deep Learning}

\author{Ruobing Han\inst{1} \and
James Demmel\inst{2} \and
Yang You\inst{3}}
\authorrunning{R. Han et al.}
%
\institute{Georgia Institute of Technology, Atlanta, GA, USA \\ \email{hanruobing@gatech.edu} \and
University of California, Berkeley, CA, USA \\ 
\email{demmel@berkeley.edu}
 \and
National University of Singapore, \\
\email{youy@comp.nus.edu.sg}}

\maketitle
\begin{abstract}
    It has been reported that the communication cost for synchronizing gradients can be a bottleneck, which limits the scalability of distributed deep learning. Using low-precision gradients is a promising technique for reducing the bandwidth requirement. In this work, we propose Auto Precision Scaling (APS), an algorithm that can improve the accuracy when we communicate gradients by low-precision floating-point values. APS can improve the accuracy for all precisions with a trivial communication cost. Our experimental results show that for many applications, APS can train state-of-the-art models by 8-bit gradients with no or only a tiny accuracy loss (<0.05\%). Furthermore, we can avoid any accuracy loss by designing a hybrid-precision technique. Finally, we propose a performance model to evaluate the proposed method. Our experimental results show that APS can get a significant speedup over state-of-the-art methods. To make it available to researchers and developers, we design and implement CPD (Customized-Precision Deep Learning) system, which can simulate the training process using an arbitrary low-precision customized floating-point format. We integrate CPD into PyTorch and make it open-source \footnote{https://github.com/drcut/CPD}.
\end{abstract}


\section{Introduction}

State-of-the-art deep learning models are becoming deeper and larger, which take an extremely long time to train.
As a result, distributed memory systems are becoming popular to train these huge models.
Most researchers are using synchronous SGD for data-parallel training \cite{goyal2017accurate,jia2018highly,you2018imagenet,ying2018image}. 
However, we can not always improve the training speed by just using more processors, as the communication cost is a non-trivial overhead for distributed systems and multi-GPU systems.
For example, communication can take 40\% of wall-clock time for BERT training on a 8 NVIDIA GTX1080Ti GPU server.
A potential solution is to use low-precision gradients \cite{jia2018highly,sun2019gradientflow}. 
However, for IEEE floating point system, previous methods can only use 16-bit for communicating gradients. 
One reason is that current communication systems only support half/single/double-precision formats. 
To solve this problem, we build a system that allows researchers to use an arbitrary low precision format (<32 bits) to communicate gradients. 
We refer to it as CPD: A High-Performance System for Customized-Precision Deep Learning. We integrate CPD into PyTorch for public usage.

We find that directly using low-precision gradients can easily hurt testing accuracy and even make the training diverge. 
One reason is that the values in gradients may easily underflow or overflow as the numerical range of the low precision is quite narrow compared to that of the high precision.
So there are lots of zeros and INF values, which can make the training process diverge. 
To solve this problem, we propose the APS (Auto-Precision-Scaling) algorithm, which is a layer-wise adaptive scheme for efficient gradients communication. 
With APS, we can make the distributed training converge with only 8 bits or even 4 bits totally for the sign, exponent (exp) and mantissa (man).
In our experiments, APS can improve the accuracy for any precision with a minor overhead. 
Compared to previous methods, the main contributions of our paper include: 
\begin{itemize}
    \item we propose APS, a layer-wise adaptive scheme, that can improve the accuracy for arbitrary low-precision formats;
    \item we are able to use several 8-bit floating point formats to train state-of-the-art classification models and segmentation models on distributed systems;
    \item we are able to use 8-bit floating point formats for gradients to train ResNet-50 on a 256-node distributed system;
    \item we build a system that can use arbitrarily customized low-precision floating-point operations and make it open-source to the public.
\end{itemize}

\section{Related Work \label{sec:related_work}}





{\bf [Gradient Sparsification]} A large DNN model typically has millions of elements in parameters and gradients. 
Researchers found that some values in gradients are much more important than others \cite{dryden2016communication}: larger values in gradients will have a greater impact on the parameter updating and the training process.
Based on this finding, some works only synchronize a part of gradients at each iteration.
There are several methods to choose the threshold and accumulate the stale gradients with new gradients \cite{sun2019gradientflow,lin2017deep,dryden2016communication}. 
For example, \cite{lin2017deep} proposed DGC, which communicates a fraction of the gradients each iteration and store the remaining gradients locally with momentum correction to maintain the accuracy. While in \cite{wangni2017gradient}, some layer's gradients will be randomly dropped out in each iteration to reduce the communication cost. All of these methods depend on gradients' magnitude rather than gradients' precision used during communication, so our method is orthogonal to these methods: we can use the above algorithms to select the gradients and use APS to communicate them with low precision.


{\bf [Gradient Quantization]} Researchers can use half-precision floating-point to communicate gradients in AlexNet/ResNet training with 100+ nodes \cite{micikevicius2017mixed, sun2019gradientflow,jia2018highly}.
The underflow/overflow issue is a serious problem in low-precision computation.
To solve this problem, \cite{micikevicius2017mixed} suggests researchers should carefully select a constant scalar to scale the loss value, which in turn will scale the gradient value. 
The constant scalars typically are different for different models and precisions.
Instead of using low-precision floating point for gradients, some researchers \cite{alistarh2017qsgd, wen2017terngrad} proposed algorithms that quantize the gradients. 
Both QSGD \cite{alistarh2017qsgd} and TernGrad \cite{wen2017terngrad} use the same idea: they encode the gradients to unbiased estimate gradients represented by fewer bits and communicate these gradients with some extra information, and finally decode these communicated results into the normal gradients. 

Although these two algorithms also use fewer bits to represent gradients, APS is significantly different from them. 
Instead of using a customized data structure to represent gradients with fewer bits, APS uses floating-point format to communicate gradients. 
APS is able to mitigate the round-off error so that we can have numerical values close to the original precision. 
APS is transparent for high level users, which means they can use the same hyper-parameters and training strategies but with less time spent on communication. 
For large scale distributed systems, it is extremely expensive to fine-tune the hyper-parameters as it will require lots of computing resources. 
Thus, it is highly necessary to maintain the same hyper-parameter set. 
Although QSGD can also maintain the hyper-parameter set, it introduces an extra hyper-parameter, the bucket size, which may significantly affect the accuracy. 
Ternary can not maintain the same hyper-parameter set because it asks users to decrease dropout ratio to keep more neurons, use smaller weight decay and disable ternarizing in the last classification layer while training on distributed systems.
Besides, compared to training on small-scale distributed systems, training on large-scale distributed systems will require lots of accumulation operations, which requires a high numerical precision. Otherwise, the results will be significantly different due to the accumulative effect. 
The validation of Ternary is only verified on small distributed systems with no more than eight nodes.
QSGD is verified on a distributed system that has only 16 nodes.
APS does not require any additional hyper-parameters, and it can maintain the hyper-parameter set used for FP32. 
Besides, we have verified the validation of APS on a large scale distributed system (256 nodes) with state-of-the-art deep learning models.
Table \ref{table:algorithm_comparing} summarizes the difference between APS and other methods.


	\begin{table}[htbp]
	\tiny
	\centering
    \begin{tabular}{|c|c|c|c|}
    \hline
    methods & same hyper & communication cost & extra hyper \\ 
     & parameter as FP32 & with gradient size L & parameter \\ 
    \hline 
    APS    & yes & allreduce(8 bits) +  allreduce(8L bits) & no    \\ 
    \hline
    loss scaling \cite{micikevicius2017mixed} & yes & allreduce(L * 16 bits)  & scaling factor   \\ 
    \hline
    TernGrad \cite{wen2017terngrad}    & no & uses special distributed system  & no                    \\ 
    \hline
    QSGD \cite{alistarh2017qsgd} & no  & depends on coding algorithm & bucket size           \\ 
    \hline
    flex16+5 \cite{koster2017flexpoint} & yes & single node. Gradients: (16L+5) bits & no \\
    \hline
    \end{tabular}
    \caption{The difference between APS and other methods.}
    \label{table:algorithm_comparing}
    \end{table}

{\bf [Low-Precision for Deep Learning]} There are several papers that explored the possibility of using a lower precision for DNN. However, most of them were focused on the inference stage. 
Recently, \cite{micikevicius2017mixed} used the half-precision format (IEEE 754 16-bit) in DNN training. With the help of loss-scaling (the scale factor is a manually-tuned hyper-parameter), they achieve a similar accuracy as the FP32 format. 
After that, \cite{wang2018training} used 8 bits in DNN training (16 bits for parts of the data) and achieve a comparable accuracy as the baseline. 
The specific design of 8 bits and 16 bits are based on the information of data distributions.
\cite{johnson2018rethinking} looked into older representations of FP to produce faster silicon.
Different from floating-point, some researchers tried using fixed-point and its variants. \cite{courbariaux2014training} used a dynamical fixed point (DFXP) format for parameters, activations and gradients. 
DFXP will change the scaling factor if overflow occurs during training. 
Instead of changing the scaling factor after overflow happens, \cite{courbariaux2014training} designed a predictor to change the scaling factor in advance to avoid overflow.
However, the previous low precision ($<$16 bits) DNN training studies are mainly focused on single node (i.e. small-batch training).
If we want to finish the training in a short time, we need distributed training on clusters.
Although some works use low precision gradient\cite{wang2018training,micikevicius2017mixed}, they do not communicate these low precision gradient. 
Low precision gradient synchronization will result in round-off error dilemma and hurt the accuracy (Sec \ref{sec:round_off_error}). In addition to saving bandwidth for synchronization, APS can be used as an algorithm to improve the accuracy for any given precision. We believe this is an important property as many new floating point formats have been proposed \cite{wang2018training}. Please see Table \ref{table:precision_range} for more details.

    \begin{figure}[htbp]
    \tiny
    \centering
    	\begin{subfigure}{.27\textwidth}
			\centering
			\includegraphics[width=\textwidth]{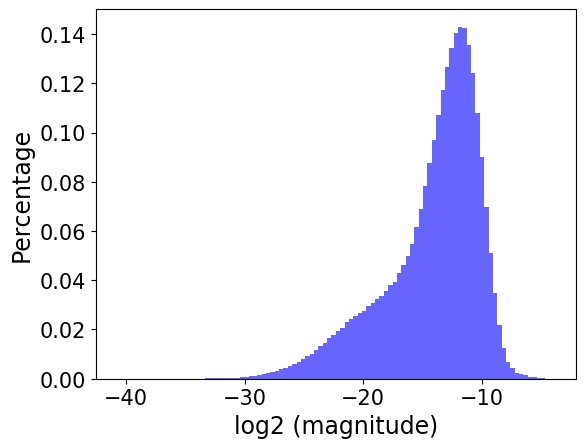}
			\caption{LeNet}
		\end{subfigure}
		 \begin{subfigure}{.27\textwidth}
			\centering
			\includegraphics[width=\textwidth]{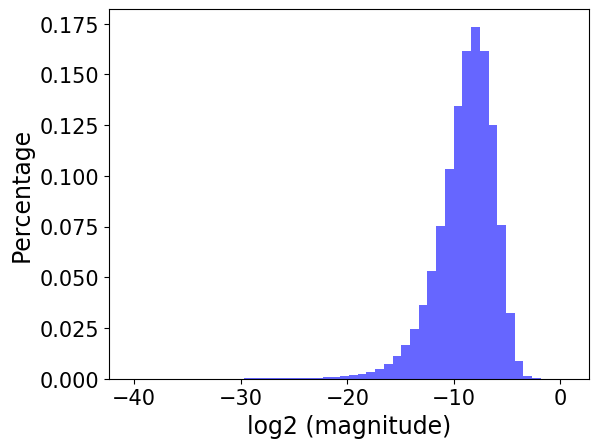}
			\caption{Resnet18}
		\end{subfigure}
		   \begin{subfigure}{.27\textwidth}
			\centering
			\includegraphics[width=\textwidth]{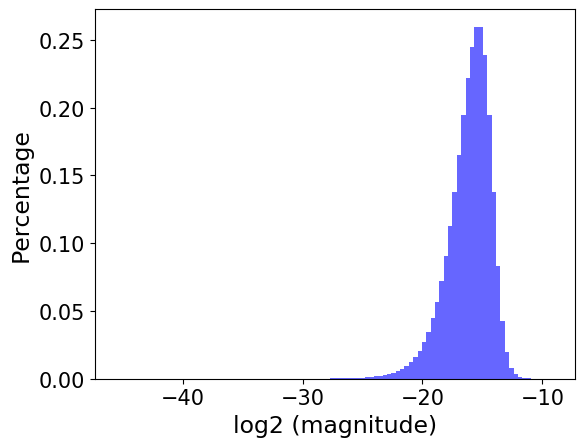}
			\caption{Resnet50}
		\end{subfigure}
	\caption{Gradients distributions for different Neural Networks. }
	\label{fig:gradient_distribution}
	\end{figure}


{\bf [Customized-Precision System]} Most state-of-the-art systems only support a fixed number of bits in a floating-point format.
For example, CUDA only supports floating-point formats with 16, 32, and 64 bits for fixed exponent/mantissa bits.
QPyTorch \cite{zhang2019qpytorch} is a recent system that allows users to assign customized number of bits to exponent/mantissa in DNN training.
However, QPyTorch has several limitations for real-world applications.
When users design a format with only a few bits for exponent, the cast results from IEEE FP32 to the low precision format are numerically incorrect, which leads to a serious bug.
Besides, it only supports IEEE 754 single-precision for all-reduce operations, which are being used at each iteration for distributed training.
To solve these problems, we develop the CPD (Customized-Precision Distributed Deep Learning) system. 

\section{APS: Auto-Precision-Scaling} \label{sec:aps_introduce}

\subsection{The limitation of the loss scaling algorithm}

\begin{figure}[!h]
    \centering
		    \begin{subfigure}{.3\textwidth}
			\centering
			\includegraphics[width=\textwidth]{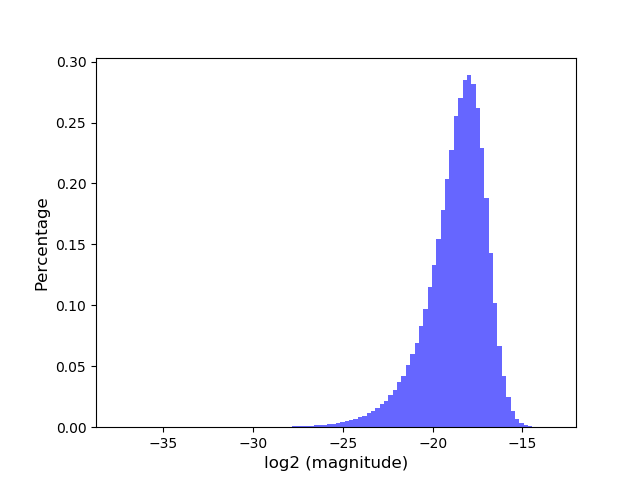}
			\caption{res5c\_2a\_weight}
		\end{subfigure}
		 \begin{subfigure}{.3\textwidth}
			\centering
			\includegraphics[width=\textwidth]{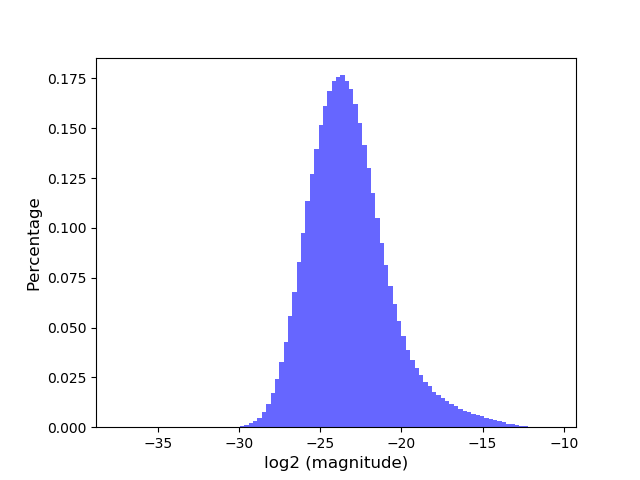}
			\caption{fc1000\_weight}
		\end{subfigure}
		    \begin{subfigure}{.3\textwidth}
			\centering
			\includegraphics[width=\textwidth]{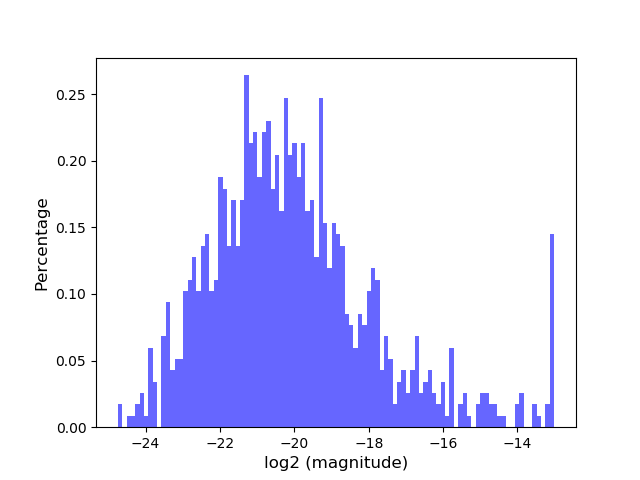}
			\caption{fc1000\_bias}
		\end{subfigure}
	\caption{Gradients distributions of different layers in ResNet50 with 8K batch size.}
	\label{fig:gradient_resnet50}
	\end{figure}

The loss scaling algorithm is being used in recent large-scale systems \cite{sun2019gradientflow,micikevicius2017mixed,jia2018highly}. 
The key idea of loss scaling is: as the ranges that can be presented by low precision and high precision are different, users can scale all layers' gradients with a factor to potentially solve the overflow/underflow problem. According to the properties of derivative, users can easily scale all gradients by multiplying the loss value with this factor (See Fig. \ref{fig:aps_comparing} (b)).
The loss scaling algorithm requires researchers to find a suitable loss scaling factor for each model, as the gradient distributions for different models are quite different in real-world applications (Fig. \ref{fig:gradient_distribution}). Besides, there are several widely used precision formats \cite{wang2018training,johnson2018rethinking}. 
For different formats, the representation ranges are also different (Table \ref{table:precision_range}). Therefore, even for the same model, the suitable loss scaling factors are different when training with different precisions. 
To make things more complicated, even within a single model, the distributions of different layers are quite different (Fig. \ref{fig:gradient_resnet50}). Previous researchers also reported the gradient distribution for a single layer also changes in training process \cite{koster2017flexpoint,courbariaux2014training}. These inconsistencies may make the loss scaling algorithm extremely unreliable in real-world applications.

    \begin{table}[htbp]
    \tiny
    \centering
    \begin{tabular}{|c|c|c|c|}
    \hline
    format     & exp bits & man bits & range \\ \hline
    IEEE 754 FP32 & 8             & 23            & [$2^{-149}, 2^{127}$]    \\ \hline
    IEEE 754 FP16 & 5             & 10            & [$2^{-24}, 2^{15}$]      \\ \hline
    BFloat16      & 8             & 7             & [$2^{-133}, 2^{127}$]    \\ \hline
    FP16 in \cite{wang2018training} & 6 & 9 & [$2^{-39}, 2^{31}$] \\ \hline
    FP8 in \cite{wang2018training} & 5 & 2 & [$2^{-16}, 2^{15}$] \\ \hline
    \end{tabular}
    \caption{Different floating-point formats have different representation ranges.}
    \label{table:precision_range}
    \end{table}

\subsection{Layer-wise precision for scaling the gradients}

To solve these problems, we propose Auto Precision Scaling algorithm (APS), which uses a layer-wise scheme to scale the gradients.
Let us refer to the layer ID as $i$ and the gradient of this layer as $grad_i$.
Then the algorithm computes the exponent values of $|grad_i|$ as $grad\_exp_i$.
Assume the model has $n$ layers, the algorithm stores a vector $E$ = $\{grad\_exp_1, grad\_exp_2, ..., grad\_exp_i, ..., grad\_exp_n\}$ in the memory and does an all-reduce operation for this vector to get the maximum value in the whole system.
Then the algorithm shifts the gradients of each layer based on the information of vector $E$ and cast them to a lower precision.
After finishing an all-reduce operation for these low precision gradients, the algorithm casts them to a higher precision and then shifts them to the original exponent. For more details, please see Algorithm \ref{alg:APS_grad}. Besides, we can synchronize the gradients for several consecutive layers as a whole tensor, which can speed up communication process by reducing the latency. 

  \begin{algorithm}[htb]
  \caption{Auto Precision scaling algorithm}
  \label{alg:APS_grad}
  \scriptsize
  \begin{algorithmic}[1]
            \Require $Gradient$: gradient (high precision)
            \Require $exp\_bit$: bits of low precision exponent
            \Require $man\_bit$: bits of low precision mantissa
            \Require $N$: numbers of distributed nodes
            
                \State $upper\_bound\_exp \gets $$2^{exp\_bits -1}$$ - 1$
                \ForAll{$g \in Gradient$}
                    \State $max\_grad\_exp \gets $ \Call{FindMaxExp}{$g *  N$}
                    
                    \State $\tilde{f} \gets $ $upper\_bound\_exp$ - \Call{AllReduce}{$max\_grad\_exp, MAX$}
                    
                    \State $g \gets g * $$2^{\tilde{f}}$$ $
                    
                    \State $low\_g \gets $\Call{Cast}{$g, exp\_bit, man\_bit$} \Comment{cast to low precision}
                    
                    \State $low\_g \gets $ \Call{AllReduce}{$low\_g, SUM$}
                    
                    \State $g \gets $\Call{Cast}{$low\_g, 8, 23$} \Comment{cast back to high precision (exp: 8, man: 23)}
                    
                    \State $g \gets g / $$2^{\tilde{f}}$$ $
                \EndFor
                
            \State
            \Function {FindMaxExp}{$Tensor$}
                \State $max\_exp \gets -INF$
                \ForAll{$i \in Tensor$}
                    \If{$i != 0$}
                        \State $tmp\_exp \gets $ \Call{ceil}{\Call{log2}{\Call{abs}{$i$}}}
                        \If{$tmp\_exp > max\_exp$}
                            \State $max\_exp \gets tmp\_exp$
                        \EndIf
                    \EndIf
                \EndFor
                \State \Return{$max\_exp$}
            \EndFunction
            \State
        \end{algorithmic}
    \end{algorithm}

\begin{figure*}[htbp]
		\begin{subfigure}{.3\textwidth}
			\centering
			\includegraphics[width=\textwidth]{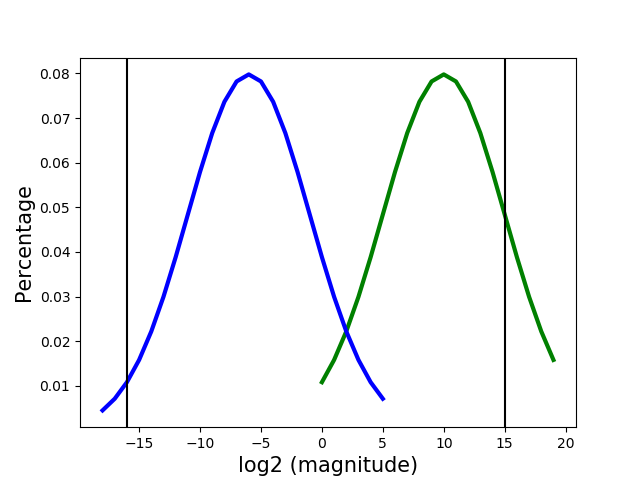}
			\caption{normal case}
		\end{subfigure}
		\begin{subfigure}{.3\textwidth}
			\centering
			\includegraphics[width=\textwidth]{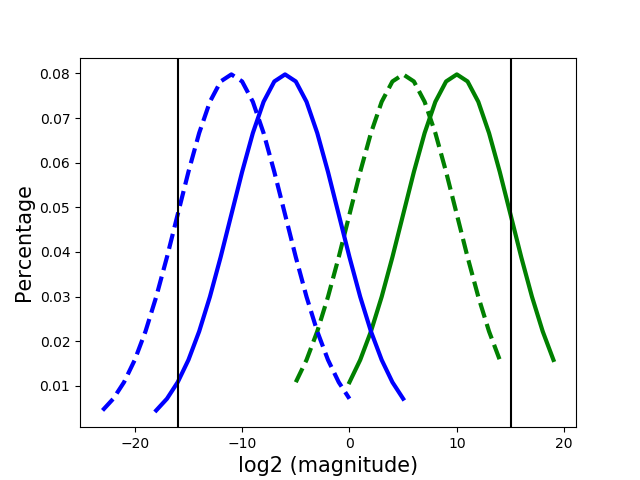}
			\caption{loss scaling}
		\end{subfigure}
		\begin{subfigure}{.3\textwidth}
			\centering
			\includegraphics[width=\textwidth]{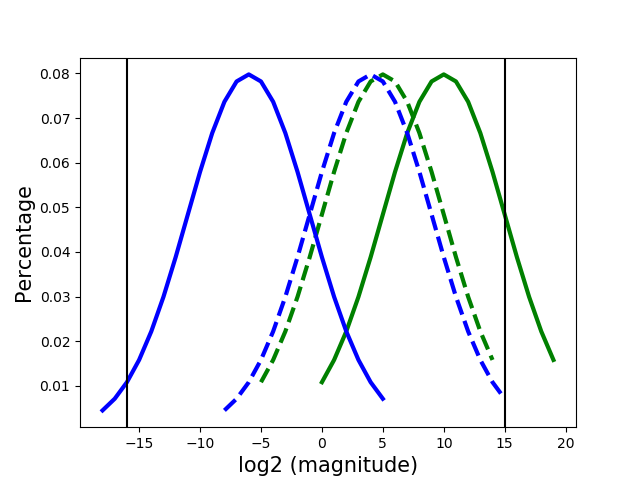}
			\caption{Auto Precision scaling}
		\end{subfigure}
		\caption{These figures show the comparison between loss scaling and APS. When we use 8 bits (exp: 5 bits, man: 2 bits), we can only represent values with exponents in [-16, 15], shown as the area between the two black lines. The gradient distributions of two layers are represented by blue/green curves separately. Values greater than $2^{15}$ will overflow and cast to INF, while values smaller than $2^{-16}$ will underflow and cast to 0. The dashed curves represent the data distributions after scaled.}
		\label{fig:aps_comparing}
	    \end{figure*}

Figure \ref{fig:aps_comparing} shows the comparison between the loss scaling algorithm and APS algorithm. When we use 8 bits (exp: 5 bits, man: 2 bits), we can only represent values with exponents in [-16, 15], shown as the area between the two black lines. 
Values greater than $2^{15}$ will overflow and cast to INF, while values smaller than $2^{-16}$ will underflow and cast to 0. The blue curve and green curve represent the gradients' distribution of two layers separately. 
The loss scaling algorithm will scale all layers' gradients with a given constant number, which is carefully selected by hand to avoid the overflow for the maximum gradients. In this case, the loss scaling algorithm will scale all gradients by $2^{-5}$. The scaled gradients are represented by dashed curves (Fig. \ref{fig:aps_comparing} (b)). Although it can avoid overflow, it will cause some small values to underflow, which will be cast to 0. 
APS algorithm will scale each layer with a different constant. In other words, the algorithm will automatically scale each layer's gradient with the greatest factor that does not cause overflow. As for the situation the figure shows, we will scale the blue layer by $2^{10}$, and the green layer by $2^{-5}$ (Fig. \ref{fig:aps_comparing} (c)).
We highlight the difference between APS and other widely-used techniques in Table \ref{table:algorithm_comparing}.

\subsection{Technical details for APS}

\subsubsection{Using the power of 2 as scaling factors} \label{sec:scaling_with_2}

\begin{figure}[htbp]
    \centering
    \includegraphics[width=75mm]{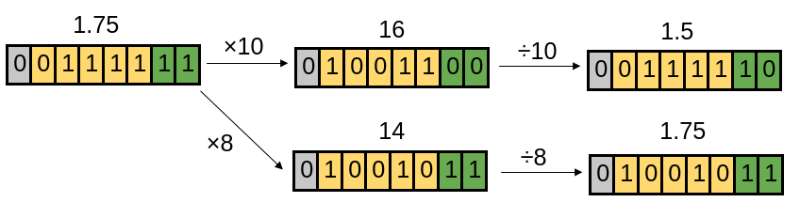}
    \caption{It is necessary to use the power of 2 as scaling factor.}
    \label{fig:precision_error}
    \end{figure}

For loss scaling \cite{micikevicius2017mixed}, users can choose arbitrary values as scaling factors. However, in APS, the algorithm will only choose a scaling factor that is the power of two. This choice can take advantage of the properties of the floating-point numbers. By doing so, we can minimize the round-off error.
For example, Fig. \ref{fig:precision_error} shows an example of using value 10 or 8 for the scaling factor. We use 8 bits precision (exp: 5 bits, man: 2 bits). The gray box denotes the sign bit, the yellow box denotes the exponent bit, and green box denotes the mantissa bit. For a normal floating-point format, when multiplied by 8 (a value that is the power of 2), only the exponent part will be changed, and the mantissa part will remain the same. So after it is multiplied and divided by 8, the output value is still the same as the input value. While using 10 as the scaling factor, both the exponent and mantissa part will be changed, which may truncate the numerical value. Either multiplied by 10 or divided by 10 will cause a round-off error.
    
\subsubsection{Trade-off between underflow and overflow}

\begin{figure}[htbp]
    \centering
    \includegraphics[width=51mm]{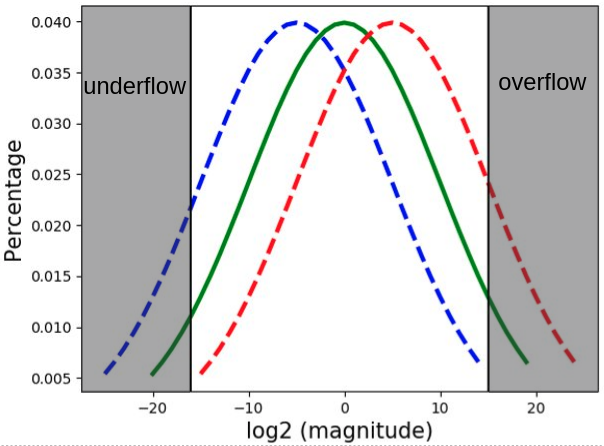}
    \caption{The green curve is the original data distribution. The blue/red dashed curves are distributions scaled by factors smaller/greater than 1.0, which leads to different underflow/overflow fractions.}
    \label{fig:trade_off}
    \end{figure}

In most cases, numbers represented by high precision formats are out of the ranges low precision formats can represent.
So the scaling technique can be a trade-off between underflow and overflow. 
An example is shown in Figure \ref{fig:trade_off}. 
The original distribution is shown in the green curve, it has both an underflow part and an overflow part. Using a scaling factor larger than 1 will move the green curve to the red curve, which is affected by overflow. In contrast, the blue curve, shifted by a scaling factor smaller than 1, is affected by underflow.
However, overflow often can be much more harmful than underflow for deep neural networks training. In backward propagation, the gradients of latter layers are used to calculate the gradients of previous layers. When the gradients of latter layers are overflow and cast to INF, all the gradients in previous layers that depend on them will also be INF. According to the rules of floating point, in most cases, the operators' outputs will be INF if there is an INF for operand. And this domino effect will make the training process diverge as we will lose lots of important information.
Therefore, our experiments and analysis indicate that we should choose a scaling factor that can avoid overflow. 
Among all these working values, we choose the largest one, which makes the smallest fraction fall into the underflow range.

\subsubsection{Find the maximum scaling factor}
The above section suggests that we should choose the maximum scaling factor which does not incur overflow. This condition is described by Equation (\ref{equation:factor_define}), we have to find the maximum value that meets this condition. In this section, we define $g$ as gradients, $f$ as the scaling factor, $\hat{p}$ as the upper bound of the required floating point precision, $N$ as the number of nodes in the system, $\hat{g}$ as the maximum element of the gradients, and $\tilde{f}$ as $\log_2factor$. Thus, we have the summation over all the distributed nodes:

\begin{equation} \label{equation:factor_define}
\footnotesize
\left| \sum_{i=1}^{N} (g_i \times f) \right| \leq \hat{p}
\end{equation}

However, as each node only knows its local gradients, it is hard to exactly get the maximum factor with negligible communication cost. So in APS, we use a heuristic algorithm to find a suitable scaling factor. We relax the bound in Equation (\ref{equation:factor_define}) as Equation (\ref{equation:relax_restriction}). 

\begin{equation} \label{equation:relax_restriction}
\footnotesize
    \left| \sum_{i=1}^{N} (g_i \times f) \right| = f \times \left| \sum_{i=1}^{N} g_i \right|\leq f \times \sum_{i=1}^{N} \left| g_i \right|\leq f \times N \times \left| \hat{g} \right|
\end{equation}

A straightforward approach is to just communicate each node's largest gradient to get the global maximum gradient and then calculate the factor. 
On top of that, we want to do further optimizations to speed up the communication process. The condition can be written as Equation (\ref{equation:new_relax_restriction}):

\begin{equation} \label{equation:new_relax_restriction}
\footnotesize
    f \leq \frac{\hat{p}}{\left| N \times \hat{g} \right|}
\end{equation}

As Section \ref{sec:scaling_with_2} suggests that we should use only the power of 2 as the scaling factor, we can further transform Equation (\ref{equation:new_relax_restriction}) ($f = 2^{\tilde{f}}$ and $\tilde{f}$ is an integer):
\begin{equation}
\footnotesize
    \begin{aligned}
    \tilde{f} < & \lceil \log_{2} (\frac{\hat{p}}{\left| N \times \hat{g} \right|}) \rceil
    = \lceil \log_{2}(\hat{p}) - \log_{2}(\left| N \times \hat{g} \right|) \rceil
    \end{aligned}
\end{equation}
So we will assign $\tilde{f} = \log_{2}(\hat{p}) - \lceil \log_{2}(\left| N \times \hat{g} \right|)  \rceil$ to meet the requirements.
For a given floating-point number, the logarithm is exactly equal to the exponent part. So instead of communicating $\left| N \times \hat{g} \right|$, we only communicate $\lceil \log_{2}(\left| N \times \hat{g} \right|) \rceil$. If we use IEEE 754 floating-point precision and communicate the former value, we have to communicate 32-bit floating point numbers. While using the latter one, we only need to communicate 8 bits, as IEEE 754 floating-point format has 8 bits for exponent.

    \begin{figure}[htbp]
        \centering
    	\begin{subfigure}{.4\textwidth}
			\includegraphics[width=50mm]{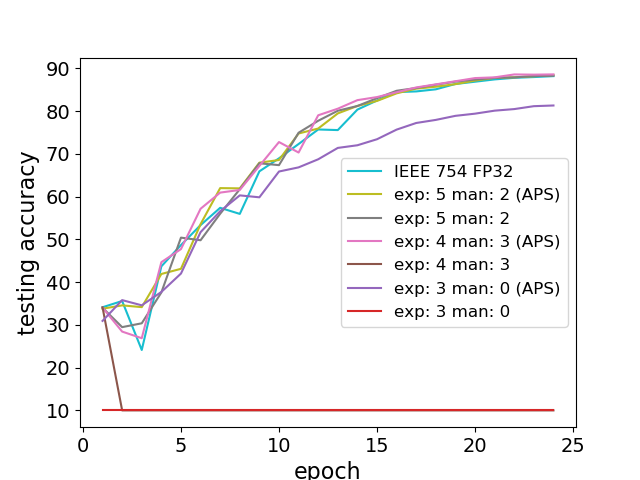}
			\caption{Davidnet}
		\end{subfigure}
		    \begin{subfigure}{.4\textwidth}
			\includegraphics[width=50mm]{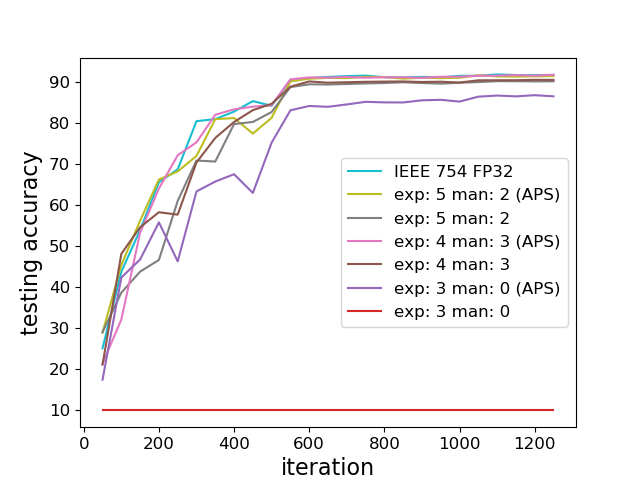}
			\caption{Resnet18}
		\end{subfigure}
	\caption{4K batch size for CIFAR10 on 8 nodes}
	\label{fig:small_scale_test}
	\end{figure}

\section{Experiments}
\label{sec:experiment}
It is hardware friendly to use a power of 2 as the number of bits.
This is efficient for both memory access and computational operations. So we tried using 4 and 8 bits for gradients in distributed training. 
We provide an emulator of CPD implementation to make sure our experiments can be reproduced on any device. 
The major concern for using mixed precision is the casting from high precision to low precision. The standard IEEE floating-point format uses the rounding-to-nearest method, while some researchers prefer stochastic rounding \cite{wang2018training,wen2017terngrad,alistarh2017qsgd} which can get an unbiased estimate for high precision values.
Although stochastic rounding has nice mathematical properties, its randomness makes it hard to reproduce. 
Also, in some situations, it is slower than the rounding-to-nearest method. 
So in the following experiments, we use round-to-nearest even method, which is a special case of the round-to-nearest method.
We fix the number of epochs as the same for all precisions for a given model.
As mentioned before, we not only focus on reducing the communication cost, but also want to make APS algorithm transparent for users, which means APS will not change the training process. Therefore, unless otherwise noted, the low precision training will use the same hyper-parameter as IEEE FP32. All hyper-parameters are referenced from previous related work \cite{davidnetpytorch, torchvision, mmsegpytorch, goyal2017accurate}, we didn’t try to fine-tune the hyper-parameters. We use IEEE FP32 for parameters and activations.

Besides, we also compare the training curves between APS and the baseline. In this way, we are able to show that using APS does not affect the training process. Most importantly, we also compare the training curves and accuracies with/without APS, to show that APS can improve the accuracy for a given precision.
In our experiments, the machines on all distributed systems run the same software environment: 64-bit Ubuntu 16.04 with CUDA toolkit 9.0, cuDNN7.6 and PyTorch1.3.1.

\subsection{Training on small-scale distributed systems \label{sec:small_scale_experiment}}    
In this section, we pick state-of-the-art deep learning models (DavidNet and Resnet18 for classification and FCN for segmentation) and train them on an 8-node distributed system, and each node has a NVIDIA V100 GPU. We use ring all-reduce \cite{gibiansky2017bringing} for all experiments in this distributed system.
For classification models, we use CIFAR10 dataset and set the batch size as 4K for both models, which means the local batch size is 512 per node in the distributed system. For ResNet18, we set the learning rate as 1.6 and use 5 epochs for learning rate warming up \cite{goyal2017accurate} from 0.1. We decay the learning rate with a factor of 0.1 at 40th and 80th epoch. We use Momentum SGD with 0.9 for $m$. 
Besides, we use a weight decay $\gamma$ of 0.0001. For DavidNet, we use Nesterov momentum with $m$ of 0.9 and set $\gamma$ as 0.256 for weight decay. We first increase the learning rate from 0 to 0.4 linearly in the first 5 epochs and then decrease it to zero linearly in the last 20 epochs.
We summarize the relationship between gradient precisions and the accuracy for DavidNet/ResNet18 in Table. \ref{table:cifar10_test}. We also show the comparison for the training curves of different precisions in Fig. \ref{fig:small_scale_test}. These results show that APS can make a significant difference in low-precision learning.

	\begin{figure}[htbp]
    \centering
    \includegraphics[width=60mm]{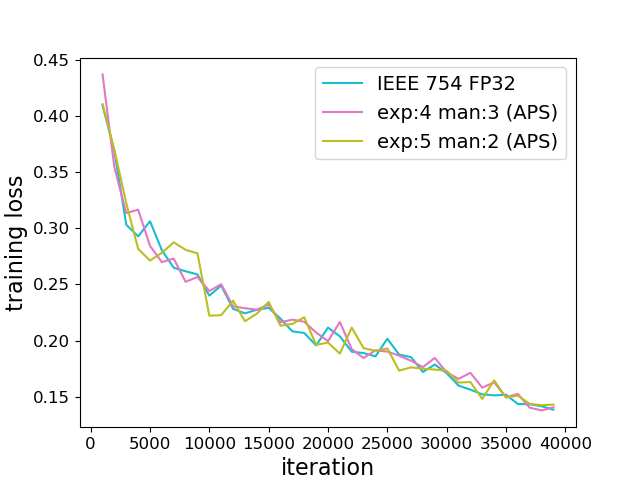}
    \caption{Training FCN on cityscapes dataset with batch size 16 on 8 nodes. Using 8 bits with APS, we can have a similar training curve as IEEE FP32 }
    \label{fig:fcn_loss}
    \end{figure}

	\begin{table}[]
	\tiny
    \begin{tabular}{|c|c|c|c|}
    \hline
    Precision (exp, man)           & Using APS & mIOU  & mAcc  \\ \hline
    (8, 23): 32bits                & /         & 75.16 & 82.84 \\ \hline
    \multirow{2}{*}{(4, 3): 8bits} & yes       & 75.88 & 84.34 \\ \cline{2-4} 
                                   & no        & 74.60 & 82.55 \\ \hline
    \multirow{2}{*}{(5, 2): 8bits} & yes       & 74.76 & 82.62 \\ \cline{2-4} 
                                   & no        & 74.41 & 82.30 \\ \hline
    \end{tabular}
    \centering
    \caption{FCN model on cityscapes with batch size 16 on 8 nodes. APS can achieve even higher accuracy with low precision (exp: 4 man:3) compared to FP32.}
    \label{table:fcn_test}
    \end{table}
	
\begin{table}[t]
\tiny
\begin{tabular}{|c|c|c|c|l}
\cline{1-4}
Model                      & Precision (exp, man)                            & Using APS & accuracy              &  \\ \cline{1-4}
                           & (8, 23): 32bits                                         & /         & 88.2                          &  \\ \cline{2-4}
                           & {\color[HTML]{000000} }                         & yes       & 88.4                          &  \\ \cline{3-4}
                           & \multirow{-2}{*}{{\color[HTML]{000000} (5, 2): 8bits}} & no        & 88.3                          &  \\ \cline{2-4}
                           & {\color[HTML]{000000} }                         & yes       & 88.6                          &  \\ \cline{3-4}
                           & \multirow{-2}{*}{{\color[HTML]{000000} (4, 3): 8bits}} & no        & 10.0                          &  \\ \cline{2-4}
                           & {\color[HTML]{000000} }                         & yes       & 81.3                          &  \\ \cline{3-4}
\multirow{-7}{*}{DavidNet} & \multirow{-2}{*}{{\color[HTML]{000000} (3, 0): 4bits}} & no        & 10.0                          &  \\ \cline{1-4}
                           & (8, 23): 32bits                                         & /         & 91.4                        &  \\ \cline{2-4}
                           &                                                 & yes       & 91.4                        &  \\ \cline{3-4}
                           & \multirow{-2}{*}{(5, 2): 8bits}                        & no        & {\color[HTML]{000000} 90.1}   &  \\ \cline{2-4}
                           &                                                 & yes       & {\color[HTML]{000000} 91.6} &  \\ \cline{3-4}
                           & \multirow{-2}{*}{(4, 3): 8bits}                        & no        & {\color[HTML]{000000} 90.4} &  \\ \cline{2-4}
                           &                                                 & yes       & {\color[HTML]{000000} 86.7}  &  \\ \cline{3-4}
\multirow{-7}{*}{ResNet18} & \multirow{-2}{*}{(3, 0): 4bits}                        & no        & 10.0                          &  \\ \cline{1-4}
\end{tabular}
\centering
\caption{Models are trained on CIFAR10 dataset with 4K batch size by 8 nodes. For all precisions, even by 4 bits, APS can make the training processes converge with little or no accuracy loss.}
\label{table:cifar10_test}
\end{table}

\begin{figure}[htbp]
    \centering
    \includegraphics[width=60mm]{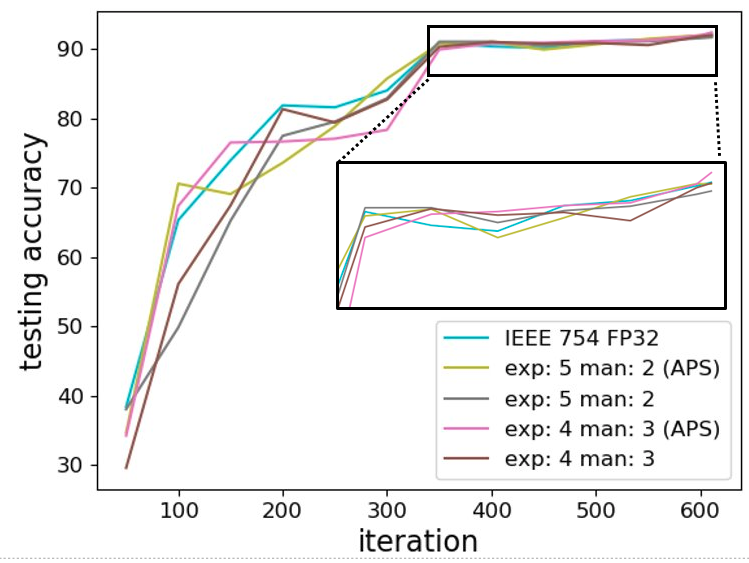}
    \caption{ResNet18 on CIFAR10 with 8K batch size by LARS. APS allows LARS to maintain the same accuracy as 32 bits while using low-precision communication.}
    \label{fig:resnet18_lars}
    \end{figure}

In addition to the classification models, we also select a state-of-the-art segmentation model, FCN \cite{Long_2015_CVPR} (with pre-trained ResNet50 for backbone), for experiments. 
We use cityscape \cite{cordts2016cityscapes} for dataset. 
We do our experiments on MMSegmentation\cite{mmsegpytorch} and use its hyper-parameter. 
In the training, we set crop size as 769$\times$769 and train 40K iterations. 
The experimental results in Table \ref{table:fcn_test} and Fig. \ref{fig:fcn_loss} show that we can use 8 bits (exp:4 man:3) to maintain the testing accuracy by APS. \footnote{\tiny mIOU: Mean Intersection-Over-Union, the average IOU over each semantic class. mAcc: (pixels in the detected area that match the ground truth)/total number of pixels in the ground truth. The higher the two metricses, the better the quality.}
LARS\cite{you2017scaling} is a state-of-the-art method being widely used for distributed training which can significantly improve the testing accuracy. 
As LARS will set the local learning rate for each layer separately based on gradients, we want to study the relationship between LARS and low-precision gradients.
We suspect LARS maybe sensitive to gradients. 
So we try using LARS with low precision gradients to see if the round-off error caused by the low precision communication hurts the accuracy or not.
We train ResNet18 on CIFAR10 dataset with 8K batch size using 8 nodes and find low precision will hurt the accuracy. On the other hand, by using APS, we can maintain the same accuracy and even improve accuracy. 
The results are shown in Table \ref{table:lars_resnet18} and Fig. \ref{fig:resnet18_lars}.
Perhaps surprisingly, the models' qualities training with low precision are close or even slightly higher than these model trained with high precision, This phenomenon is also reported in \cite{lin2017deep, micikevicius2017mixed}. This may be due to the fact that low precision can relieve overfitting, like L1 normalization.

\begin{table}[tbp]
\tiny
\centering
\begin{tabular}{cccll}
\cline{1-3}
\multicolumn{1}{|c|}{Precision (exp, man)}    & \multicolumn{1}{c|}{Using APS} & \multicolumn{1}{c|}{testing accuracy}              &  &  \\ \cline{1-3}
\multicolumn{1}{|c|}{(8, 23): 32bits}                 & \multicolumn{1}{c|}{/}         & \multicolumn{1}{c|}{92.072}                        &  &  \\ \cline{1-3}
\multicolumn{1}{|c|}{}                        & \multicolumn{1}{c|}{yes}       & \multicolumn{1}{c|}{{\color[HTML]{000000} 92.44}}  &  &  \\ \cline{2-3}
\multicolumn{1}{|c|}{\multirow{-2}{*}{(4,3): 8bits}} & \multicolumn{1}{c|}{no}        & \multicolumn{1}{c|}{{\color[HTML]{000000} 92.036}} &  &  \\ \cline{1-3}
\multicolumn{1}{|c|}{}                        & \multicolumn{1}{c|}{yes}       & \multicolumn{1}{c|}{{\color[HTML]{000000} 92.015}} &  &  \\ \cline{2-3}
\multicolumn{1}{|c|}{\multirow{-2}{*}{(5,2): 8bits}} & \multicolumn{1}{c|}{no}        & \multicolumn{1}{c|}{{\color[HTML]{000000} 91.737}} &  &  \\ \cline{1-3}
\multicolumn{1}{l}{}                          & \multicolumn{1}{l}{}           & \multicolumn{1}{l}{}                               &  & 
\end{tabular}
\caption{ResNet18 with LARS. APS can improve the accuracy for both (exp:5, man:2) and (exp:4, man:3). It can even get a higher accuracy than 32-bit precision.}
\label{table:lars_resnet18}
\end{table}

\subsection{Training on large-scale distributed systems}

We train ResNet50 \cite{he2016deep} on a 256-node distributed system. Instead of ring all-reduce used in Section \ref{sec:small_scale_experiment}, we use the Hierarchical all-reduce \cite{jia2018highly, sun2019gradientflow}: we partition the nodes into 16 groups, and assign a $master$ node for each group. Each all-reduce operation will finish 3 steps: (1) within each group, all worker nodes send their local gradients to the master node; (2) we conduct the ring all-reduce across all the master nodes; (3) within each group, the master node broadcasts the global gradients to all the worker nodes.
There are two reasons why we use the hierarchical all-reduce approach:

    \begin{figure}[t]
    \centering
    \includegraphics[width=45mm]{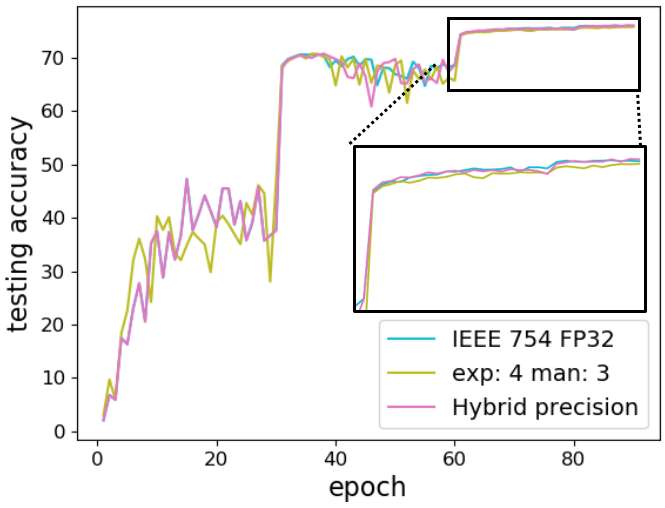}
    \caption{Training ResNet50 with 32 bits, 8 bits, and hybrid precision.}
    \label{fig:resnet50_aps}
    \end{figure}   
    
    
    \begin{figure}[t]
    \centering
    \includegraphics[width=66mm]{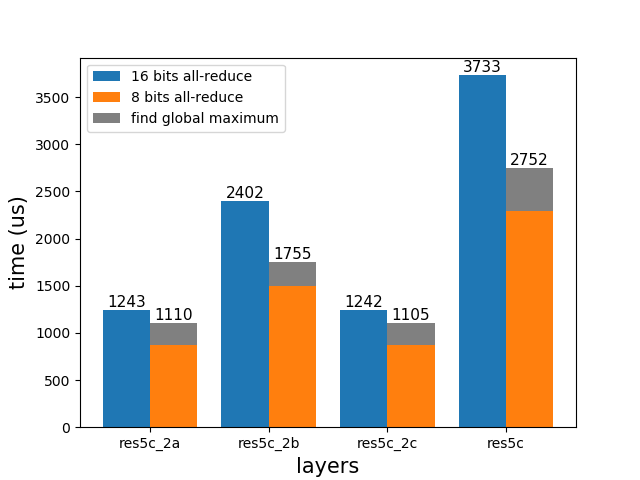}
    \caption{The time of communication on 32 nodes.}
    \label{fig:resnet50_performance}
    \end{figure}   

\begin{itemize}
\item \textbf{Performance}: the ring all-reduce with $p$ nodes need to finish $2(p-1)$ steps (each step transfers the same amount of data). 
The hierarchical all-reduce with a group size of $k$ only needs $4(k-1)+2(p/k-1)$ steps. In our experiments with 256 nodes and a group size of 16, we only need to finish 74 steps, instead of 510 steps for using ring all-reduce.

\item \textbf{Round-off error \label{sec:round_off_error}}: when we use a low precision floating point to add a small number with a large number, the smaller number may be truncated and cast as zero in this addition operation. 
This situation is common in all-reduce process. 
To avoid this problem, we should try to minimize the number of large-and-small additions.
If we use ring all-reduce, we have to add a local gradient with the summation of all other nodes' local gradients in the last step. 
The summation may be 255x larger than this local gradient if we have 256 nodes.
When we use the hierarchical all-reduce, we have only 16 nodes for intra-group reduction. 
In this situation, the last step will add a local gradient with a 15x larger gradient.
The situation is the same as inter-group all-reduce among 16 master nodes.
\end{itemize}

Taking the above two factors into account, we choose to use the hierarchical all-reduce with a group size of 16. In the following section, we prove that this group size can minimize the round-off error.
We use 8K batch size to train ResNet50 on ImageNet dataset \cite{deng2009imagenet}. 
As APS does not require us to modify the hyper-parameters, we use the same setting and data preprocessing as \cite{goyal2017accurate}, except the learnable scaling coefficient $\gamma$ is initialized as 1 for all BN layers in our experiments.
We adopt the initialization of \cite{He_2015_ICCV} for the 1000-way fully-connected layer.
Based on the suggestions of \cite{wang2018training,wen2017terngrad}, we use IEEE FP32 for the gradient of the last layer (i.e. classification layer) and low precision for all other layers' gradient. We also have the experimental results of using low precision for all layers, please see Fig. \ref{fig:resnet50_aps} for details.
We try using the APS algorithm on different precisions (Table \ref{table:APS_resnet_precision}), and find APS only needs 8 bits to achieve roughly the same accuracy as the standard 32-bit format. 
We can further improve the accuracy by hybrid precision: using FP32 for the first 30 epochs and 8 bits for the last 60 epochs. 
This method can help us maintain the same accuracy as IEEE FP32 (Table \ref{table:last_layer_precision}).

    
\begin{table}[]
\tiny
\begin{tabular}{|c|c|c|c}
\cline{1-3}
Precision (exp, man) & with APS       & top-1 accuracy &                      \\ \cline{1-3}
(8, 23): 32bits                & / & 76.02          &                      \\ \cline{1-3}
\multirow{2}{*}{(5,2): 8bits} & yes            & 75.98          &                      \\ \cline{2-3}
                     & no             & 71.00          &                      \\ \cline{1-3}
\multirow{2}{*}{(4,3): 8bits} & yes            & 75.93          &                      \\ \cline{2-3}
                     & no             & 0.1            &                      \\ \cline{1-3}
(8, 23) + (4, 3)     & yes            & 76.09          & \multicolumn{1}{l}{} \\ \cline{1-3}
\end{tabular}
\centering
\caption{We use APS to train ResNet50 with 8K batch size (256 nodes). APS can improve the accuracy for 8-bit gradient. With APS, we can use 8-bit gradient for the whole training process, with only a tiny loss in testing accuracy (<0.05\%). }
\label{table:APS_resnet_precision}
\end{table}


    \begin{table}[]
    \tiny
    \begin{tabular}{|c|c|c|}
    \hline
    Precision for other layers & Precision for the last classification layer & top-1 accuracy \\ \hline
    \multirow{2}{*}{(5, 2)}    & (5,2)                                       & 75.08          \\ \cline{2-3} 
                               & FP32                                        & 75.98          \\ \hline
    \multirow{2}{*}{(4, 3)}    & (4, 3)                                      & 75.46          \\ \cline{2-3} 
                               & FP32                                        & 75.93          \\ \hline
    \end{tabular}
    \centering
    \caption{Training ResNet50 with APS by low precision for different layers. We can improve the accuracy by using high precision for the last classification layer.}
    \label{table:last_layer_precision}
    \end{table}

We also have a comparison between different group sizes and present the results in Table \ref{table:resnet50_group_size}. 
To further explain the difference between different group sizes, we compare the average round-off error for the gradient of the first convolutional layer's weight using 8 bits (exp: 5, man: 2) and present the result in Table \ref{table:group_size_error}. 
The average round-off error is described by Equation \ref{equation:round_off_error} (the gradients got by the high/low precision communication are denoted as $grad\_h$ and $grad\_l$ separately and we assume there are $N$ elements in the gradient tensor). 
It shows using the hierarchical all-reduce can decrease the round-off error compared to ring-allreduce.
It also shows that 16 is the most suitable group size for a 256-node distributed system.

\begin{equation} \label{equation:round_off_error}
\footnotesize
    average\_round\_off\_error = \frac{\sum_{i=0}^{N} \left | \frac{grad\_h_i - grad\_l_i}{grad\_h_i}  \right |}{N}
\end{equation}

\begin{table}[]
\tiny
\begin{tabular}{|c|c|c|}
\hline
Precision (exp, man)                & group size & top-1 accuracy             \\ \hline
                         & 32         & {\color[HTML]{000000} 74.95} \\ \cline{2-3} 
\multirow{-2}{*}{(4, 3): 8bits} & 16         & {\color[HTML]{000000} 75.46} \\ \hline
                         & 32         & {\color[HTML]{000000} 74.91} \\ \cline{2-3} 
\multirow{-2}{*}{(5, 2): 8bits}  & 16         & {\color[HTML]{000000} 75.08} \\ \hline
\end{tabular}
\centering
\caption{For ResNet-50 on a 256-node system, using a group size of 16 can improve the accuracy compared to a group size of 32 for 8 bits, as it can reduce the round-off error. We use low precision gradients for all layers}
\label{table:resnet50_group_size}
\end{table}

\begin{table}[]
\centering
\tiny
\begin{tabular}{|c|c|c|c|c|c|c|l}
\cline{1-7}
group size & 4 & 8 & 16 & 32 & 64 & 256 (ring all reduce)                      \\ 
\cline{1-7}
round-off error & 55\% & 44.21\% & 41.83\% & 49.62\% & 58.21\% & 85.22\%                       \\ 
\cline{1-7}
\end{tabular}
\caption{The average round-off error for the first convolutional layer's weight of ResNet-50 using 8 bits (exponent: 5 bits, mantissa: 2 bits) on a 256-node cluster.}
\label{table:group_size_error}
\end{table}

\subsection{Performance Analysis\label{sec:performance}}
In this section, we analyze the performance of APS on a distributed system with 32 V100 GPUs. In detail, there are 4 servers, each servers is installed with 8 GPUs. All servers are connected by InfiniBand and share a distributed file system.

{\bf [Layer-wise performance analysis]} Figure \ref{fig:resnet50_performance} shows the time cost for synchronizing gradients of some layers in ResNet50 (gradient shape of each layer: res5c\_branch2a: 2048*512, res5c\_branch2b: 512*512*3*3, 
\newline 
res5c\_branch2c: 512*2048).
The blue bars denote the time cost by using half precision without APS. The bars on the right set of each blue bars show the total time for using APS to communicate the same gradient. The gray bars denote the time cost to get the global maximum gradient. The orange bars denote the time cost to communicate gradients using 8 bits. For all layers, APS with 8 bits can speed up the communication process.
Our experiments show that merging short messages into a single one can reduce the overall communication time. Here, res5c\_2a, res5c\_2b, res5c\_2c are three consecutive layers in ResNet50. We synchronize them as a whole, and present the result on the rightmost column in Figure \ref{fig:resnet50_performance}. We can achieve a 1.36$\times$ speedup over half-precision.

\begin{figure}[htbp]
    \centering
    	\begin{subfigure}{.3\textwidth}
			\centering
			\includegraphics[width=\textwidth]{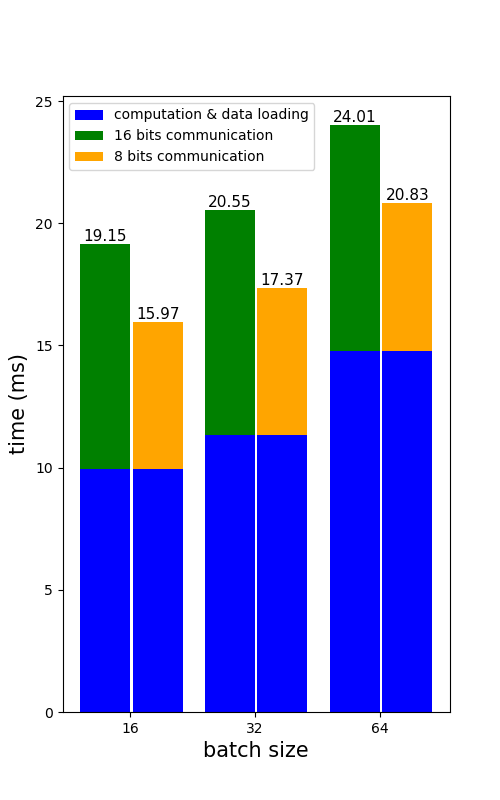}
			\caption{DavidNet}
		\end{subfigure}
		\begin{subfigure}{.3\textwidth}
			\centering
			\includegraphics[width=\textwidth]{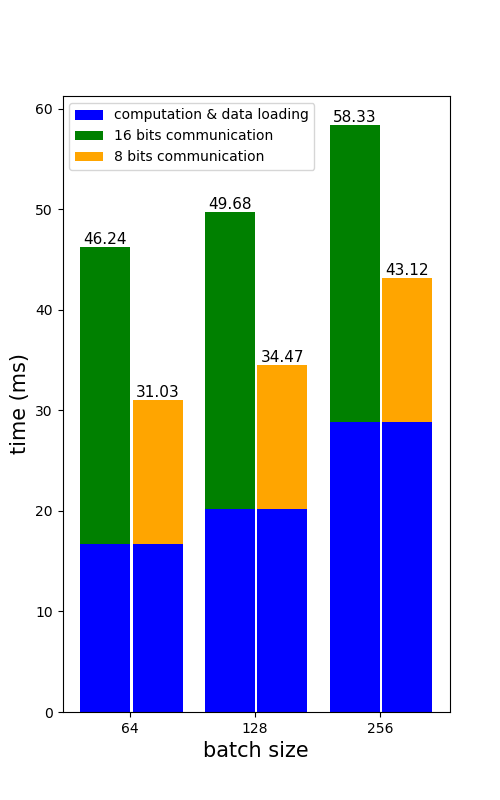}
			\caption{ResNet18}
		\end{subfigure}
    	\begin{subfigure}{.3\textwidth}
			\centering
			\includegraphics[width=\textwidth]{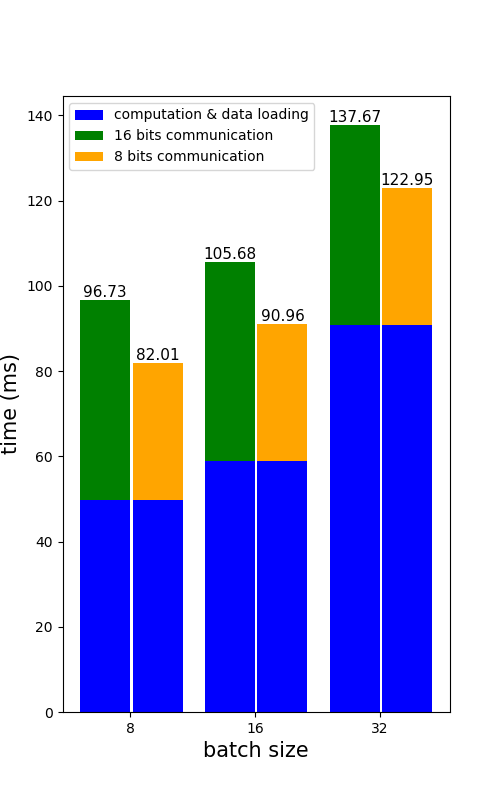}
			\caption{ResNet50}
		\end{subfigure}
	\caption{Time cost per iteration (8 bits by APS is consistently faster). }
	\label{fig:end2end_performance}
	\end{figure}

    {\bf [End-to-End performance analysis]} We also analyze the end to end performance for using 8 bits gradients with APS algorithm. We record the time for each iteration. Half-precision training \cite{jia2018highly, sun2019gradientflow}(using half-precision for both computation and communication) is used for baseline. The baseline’s time (t\_base) includes two parts: (1) the gradients all-reduce operation and (2) the computation and data processing time. APS’s communication time (t\_aps) includes the time for three parts: (1) gradients all-reduce operation, (2) reducing the global maximum value for each gradient, and (3) the computation and data processing time. In Fig. \ref{fig:end2end_performance}, we show the time cost for iterations with different batch size for different models. For ResNet-50, with batch size 32, t\_base = 137.67ms and t\_aps = 122.95ms per iteration. The speedup is 1.12x. While for batch size 8, the speedup is 1.18x. For ResNet-18 with batch size 256, the speedup is 1.35x. Even for DavidNet that has relative few parameters, we can also achieve 1.15x speed up with batch size 16. Considering ResNet and DavidNet have been well optimized by industry vendors, we think this level of speedup is very good. As the communication cost will increase with a larger number of nodes, we believe APS can achieve a higher speedup for larger systems. We are also designing a new computer architecture with a startup company for APS. We believe APS will get a higher speedup on specialized hardware. The speedup will also be higher in the federated learning situation where the computation is done on mobile devices and the communication is conducted over the internet.

\section{CPD: Customized-Precision Deep Learning \label{sec:CPD}}

Because of the limitations in QPyTorch, we built CPD to emulate the low-precision training for our experiments. 
CPD has the following functions, which are not supported by any previous systems: (1) arbitrary low-precision, with number of exponent bits $<=8$ and number of mantissa bits $<=$23; (2) various accumulation strategies (e.g. Kahan summation algorithm \cite{higham2002accuracy}); (3) using any low-precision accumulator for GEMM and reduce/all-reduce function.

\begin{figure}[htbp]
    \centering
    	\begin{subfigure}{.45\textwidth}
			\centering
			\includegraphics[width=\textwidth]{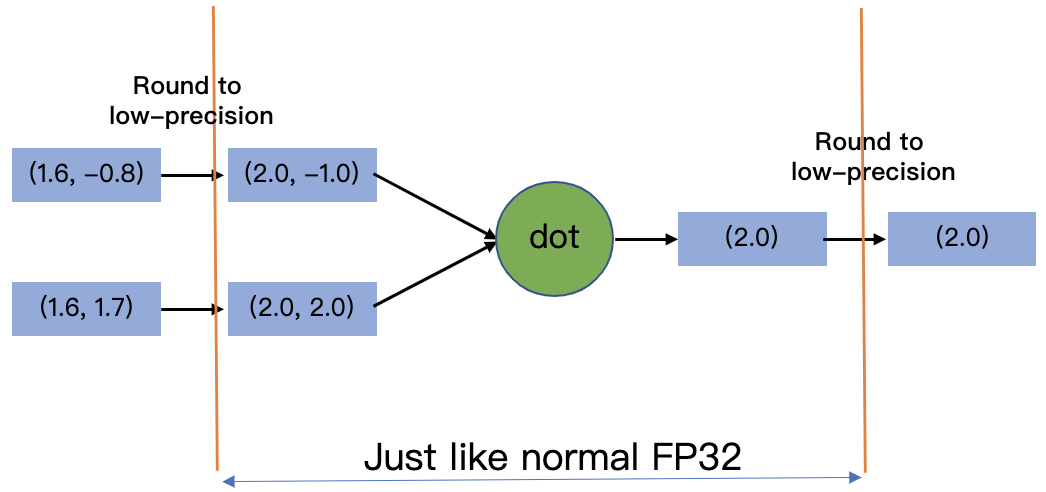}
			\caption{QPyTorch}
		\end{subfigure}
		    \begin{subfigure}{.45\textwidth}
			\centering
			\includegraphics[width=\textwidth]{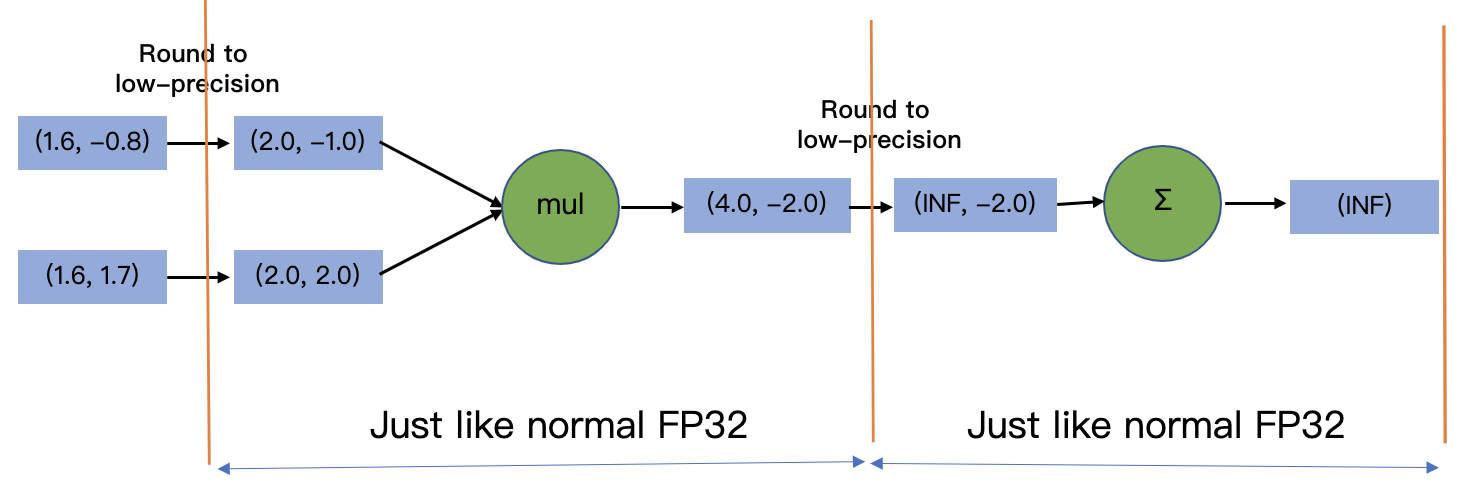}
			\caption{CPD}
		\end{subfigure}
	\caption{3-bit floating point (2-bit exp) for vector multiplication.}
	\label{fig:bugs}
	\end{figure}

Accumulation is a common operation in Deep Learning (e.g. GEMM and all-reduce). During accumulation, a number maybe added by a much larger number.
In this case, the smaller number may be truncated to zero and does not affect the accumulation.
To avoid losing the information of small values during accumulation, we can use a higher precision to store the accumulator.
Although higher precision accumulators can maintain accuracy, they will cost more energy and hurt the performance.
So the precision of the accumulator is a key factor for low precision computation.
To the best of our knowledge, existing systems only support using IEEE FP32 for the accumulator, while CPD allows users to apply arbitrary low precision ($\leq$32 bits) for the accumulator. This feature is significantly helpful for hardware designers. An example of 3-bit accumulator is shown in Fig. \ref{fig:bugs}.
    
We can also use different accumulation strategies to maintain accuracy.
CPD supports not only default sequential summation, but also the Kahan summation algorithm. It also allows users to implement their own strategies.
 
\section{Conclusion\label{sec:conclusion}}
Auto Precision Scaling (APS) is a flexible low-precision technique that can reduce the communication cost. It can train several state-of-the-art applications by 8 bits for gradient communication without losing accuracy.
APS can save the bandwidth and improve the accuracy for any given low-precision with almost no cost. 
For low-precision formats in our experiments, APS can improve the accuracy for all of them.
Besides, we can train ResNet-50 by a hybrid precision to get the same accuracy as the baseline with the same number of epochs on 256 nodes.
We analyze the time of APS for gradient communication and find that the saving in time is larger than the additional cost in compute.
Furthermore, we built the CPD system that allows users to simulate any arbitrary low-precision format. 
We integrate CPD into PyTorch and make it open source to the public.


\bibliographystyle{splncs04}
\bibliography{references}

\end{document}